\documentclass[sigconf]{acmart}
\usepackage{graphicx}
\usepackage{wasysym}
\usepackage{tcolorbox}
\usepackage{tabularx}
\usepackage{subcaption}
\usepackage{framed}
\usepackage[nolist,nohyperlinks]{acronym}

\acmConference[CPSIoTSec]{The 6th Joint Workshop on CPS \& IoT Security and Privacy}{October 18th, 2024}{Salt Lake City, UT, USA}
\title{RTFM}
\subtitle{How hard are IoT platform providers making it for their developers?}
\author{Andrew Baldrian}
\affiliation{
  \institution{University of Bristol}
  \city{Bristol}
  \country{UK}
}
\email{andy.baldrian@bristol.co.uk}
\author{Joseph Hallett}
\affiliation{
  \institution{University of Bristol}
  \city{Bristol}
  \country{UK}
}
\email{joseph.hallett@bristol.ac.uk}

\DeclareFontFamily{U}{wasy}{}
\DeclareFontShape{U}{wasy}{m}{n}{
     <-5.5> wasy5
  <5.5-6.5> wasy6
  <6.5-7.5> wasy7
  <7.5-8.5> wasy8
  <8.5-9.5> wasy9
     <9.5-> wasy10
}{}
\DeclareFontShape{U}{wasy}{b}{n}{
 <-10> ssub * wasy/m/n
 <10-> wasyb10
 }{}
\DeclareFontShape{U}{wasy}{bx}{n}{ <-> ssub * wasy/b/n}{}
\DeclareFontShape{U}{wasy}{m}{sl}{ <-> wasysl10 }{}
\DeclareFontShape{U}{wasy}{m}{it}{ <-> ssub * wasy/m/sl }{}

\acrodef{API}{Application Programming Interface}
\acrodef{CA}{Certificate Authority}
\acrodef{EU}{European Union}
\acrodef{IDE}{Integrated Development Environment}
\acrodef{IDS}{Intrusion Detection System}
\acrodef{IoT}{Internet of Things}
\acrodef{IIoT}{Industrial Internet of Things}
\acrodef{NDA}{Non-Disclosure Agreement}
\acrodef{OS}{Operating System}
\acrodef{PSA}{Platform Security Architecture}
\acrodef{PCB}{Printed Circuit Board}
\acrodef{PUF}{Physically Unclonable Function}
\acrodef{PKI}{Public Key Infrastructure}
\acrodef{TEE}{Trusted Execution Environment}
\acrodef{SoC}{System-on-a-Chip}
\acrodef{UK}{United Kingdom}
\acrodef{USA}{United Stats of America}

\begin{abstract}
  \ac{IoT} devices routinely have security issues, but are the platform designers providing enough support to \ac{IoT} developers for them to easily implement security features for their platforms?  We surveyed the documentation, code and guidance from nine \ac{IoT} manufacturers to look at what guidance they provided for implementing three security features required by several security standards (secure boot, device identity keys and unique per device passwords).  We find that more needs to be done to support developers if we want them to adopt security features---especially in the face of incoming legislation that will require developers to implement them.
\end{abstract}

\begin{document}
\maketitle

\section{Introduction}
\ac{IoT} devices are everywhere, but have a reputation for being insecure~\cite{barcena2015insecurity}.
Providing usable documentation and example code is known to help developers implement security features~\cite{green2016developers,patnaik_usability_2019,chowdhury_developers_2021}; so to explore why \ac{IoT} devices might be insecure we looked at the resources \ac{IoT} platforms provide to help developers implement security features. 

Internet connected devices are expected to exceed 29 billion units by the end of 2030~\cite{statista_iot_2024}. These \ac{IoT} devices are used for a multitude of applications: for example, roadside weather monitoring, video doorbells, and internet-enabled fridge-freezers~\cite{asghari_internet_2019,khanna_internet_2020}.
Concerns have been raised around the security of \ac{IoT} devices. In 2016, the Mirai botnet infected over 600 thousand devices simply by exploiting common default passwords used by \ac{IoT} manufacturers~\cite{antonakakis_understanding_2017}.  Increasingly these devices are seen as being vulnerable~\cite{neshenko_demystifying_2019}.
Multiple organisations have publish recommendations or standards to minimise these vulnerabilities~\cite{chen_unified_2023}.
As the use of this technology grows, the world of cyber physical systems and \ac{IoT} is also interacting with the introduction of the \ac{IIoT}~\cite{sisinni_industrial_2018}; further increasing the reach of these devices.

The explosion in the number of \ac{IoT} devices and the concerns around their possible security vulnerabilities has fueled the need for regulation. For example, the \ac{UK} Product Security and Telecommunications Infrastructure (Product Security) regime~\cite{uk_uk_2024}, or the \ac{EU} Cyber Resilience Act (EU-CRA)\cite{eu_cyber_2023}.
These regulations attempt to address these security and privacy concerns by defining a set of minimal requirements that an \ac{IoT} device should conform to~\cite{chiara_iot_2022}.
While the regulations may be welcome by industry bodies and governments~\cite{iotf_uk_2021}, they may also increase the cost of developing these devices with a possible knock-on effect of increasing the costs to the consumer~\cite{mcdonald_how_2022}.
Engineers building these devices must address these requirements or \ac{IoT} device manufacturers may lose access to geographical markets.

\ac{IoT} devices are built using existing processors, communication peripherals, sensors and actuators. 
Chip manufacturers have recognised the need for an \ac{IoT} platform that can be used by \ac{IoT} device manufacturers to construct these \ac{IoT} devices.
These platform manufacturers deliver \ac{IoT} platforms with three core components: processors, memory and communications.
To achieve the regulatory goals, device engineers need \ac{IoT} platforms to also provide a set of security features and the means to develop solutions that meet the requirements of the regulations.
This paper reviews some of the leading \ac{IoT} platform manufacturers, looking at the security features they provide and the documentation available to engineers and the supporting material such as code examples, to help them meet these goals;
focusing on three security features, necessary to implement many of the regulatory goals:
\begin{itemize}
  \item Secure boot
  \item Device identity key
  \item Unique per device password.
\end{itemize}
This paper asks the following research questions:
\begin{description}
  \item[RQ1.] To what extent are \ac{IoT} platforms providing hardware functionality for device engineers to implement basic \ac{IoT} security tasks? 
  \item[RQ2.] How are these platform manufacturers supporting device engineers to use these features correctly?
  \item[RQ3.] What additional support is provided to help device engineers take advantage of security features to meet their goals?
\end{description}
We find that whilst secure boot support is relatively common, functionality for other features is hidden (RQ1) and that in all cases documentation and code examples (if any are given) are relatively poor (RQ2). Cloud services or third party software my be used to provide basic security features but do not meet many of the additional regulatory requirements (RQ3).
This suggests that more needs to be done to support developers if we want to increase the adoption of security features.
\section{Standards and security recommendations}
\begin{table*}
  \setlength{\extrarowheight}{.3\baselineskip}
  \begin{tabularx}{\linewidth}{|l|X|X|X|}
    \hline
    Standard & Secure Boot & Device Identity Keys & Unique Per Device Passwords \\
    \hline
    CSA~\cite{csa_future_2016}
      & 11 Design Secure Bootstrap Functions
      &
      &
    \\
    CSA IWG~\cite{csa_identity_2015}
      &
      & 08 Consider design updates to your \ac{PKI} environment to support provisioning of certificates to \ac{IoT} devices in your organization.
      & 02 Do not deploy \ac{IoT} resources without changing default passwords for administrative access.
    \\
    CSDE~\cite{csde_c2_2021}
      & B.2 Hardware Rooted Security
      & 5.1.1 Device Identifiers
      & 5.1.2 Secured Access
    \\
    DSIT~\cite{dsit_code_2018} 
      &
      &
      & 1. No default passwords
    \\
    ENISA~\cite{enisa_baseline_2017}
      & GP-TM-02, GP-TM-03
      & GP-PS-10
      & GP-TM-09, GP-TM-22, GP-TM-23
    \\
    ETSI EN 303 645~\cite{etsi_cyber_2020} 
      & Provision 5.7-1
      & Provision 5.4-2
      & Provision 5.1-1, Provision 5.1-2
    \\
    GSMA CLP13 V2.2~\cite{gsma_iot_2020} 
      & 6.17.1 Risk
      & 6.20 Authenticating an Endpoint Identity
      &
    \\
    IEC 6244-4-2
      & CR 7.4, CR 7.6
      & CR 1.2
      &
    \\
    ICC~\cite{schrecker_security_2026}
      & 7.3 Endpoint Protection, 8.7.1 Boot Process Integrity
      & 8.5 Endpoint identity, 11.7 Identity management
      &
    \\
    IoTSF~\cite{iotsf_iotsf_2021}
      & 2.4.4.1-4
      & 2.4.8.1
      & 2.4.8.3,  2.4.8.13, 2.4.10.4, 2.4.11.1
    \\
    NIST IR 8259A~\cite{fagan_iot_2020} 
      & Device Configuration
      & Device Identification
      &
    \\
    OCF~\cite{ocf_ocf_2020}
      & 14.3
      &
      &
    \\
    PSA~\cite{psacertified_psa_2023}
      & C1.2, D1.1
      & C1.4, S2.1, D4,6
      & S5.1, S5.2, D4.1, D4.5
    \\ 
    \hline
  \end{tabularx}
  \caption{\ac{IoT} standards reviewed as part of this study highlighting where each of the security features examined in this study (secure boot, device identity keys and unique per device passwords) are discussed within.}
  \label{tab:standards}
\end{table*}

The diversity of \ac{IoT} devices and their development in multiple markets results in multiple standards~\cite{chen_unified_2023} covering multiple different markets, for example with medical devices, the Hippocratic Oath for Connected Medical Devices~\cite{iamthecavalry_hippocratic_2016} or for telecommunications the GSMA CLP.13 - \ac{IoT} Security Guidelines Endpoint Ecosystem~\cite{gsma_iot_2020}.
Standards such as ETSI EN 303 645 Cyber Security for Consumer Internet of Things: Baseline Requirements~\cite{etsi_cyber_2020}, or NIST IR 8259A Core Device Cybersecurity Capability Baseline~\cite{fagan_iot_2020} provide a technical baseline set of requirements for device engineers to follow.
Often these standards will reference related standards that focus on information processing or privacy, for example ISO/IEC 27002 Information security, cybersecurity and privacy protection — Information security controls~\cite{1400-1700_isoiec_2022}.
While these standards may not be directly focused on \ac{IoT} devices, they do provide requirements for data processing and data privacy.
This results in a complex set of security and privacy standards for device engineers to navigate.

To help device engineers manage these different sets of requirements, a number of organisations have published security frameworks.
For example, the \ac{IoT} Security Foundation provide Secure Design Best Practice Guides and an \ac{IoT} Security Assurance Framework~\cite{iotsf_iotsf_2021}, providing a process that device engineers can follow to enable them to build compliant devices.
\ac{PSA} Certified provides a certification program and security framework for device engineers to follow~\cite{psacertified_psa_2023}.
The GSMA has an \ac{IoT} Security Assessment with a checklist~\cite{gsma_iot_2024} of different requirements that device engineers should consider when designing an \ac{IoT} device.

From a review of 13 IoT standards, we have selected three security features mentioned in multiple standards~(Table~\ref{tab:standards}). These  security features are not sufficient for an \ac{IoT} deployment, but serve as a indication of the abilities of the \ac{IoT} platform.
The security features are selected to highlight the differing types of feature that device engineers may face when implementing an \ac{IoT} device.
The security features may also depend on specific hardware functionality that must be available in the \ac{IoT} platform to implements this security feature.

\subsection{Secure boot}

The boot process is the act of initialising the hardware and loading the firmware into memory when the \ac{IoT} device is turned on.
The notion of a secure boot process requires that the firmware can be cryptographically validated as part of a \textit{root of trust}~\cite{regenscheid_bios_2011} process.
This is normally implemented using a \ac{TEE}, such as the Arm TrustZone~\cite{arm_trustzone_nodate}.
The boot sequence can be a single or two-stage process:

\subsubsection{Load the boot loader}
The processor has a built-in immutable boot loader; its job is to validate the signature of the boot loader and load the second stage boot loader into memory.

\subsubsection{Load the firmware}
The second stage boot loader will then validate the signature on the firmware application, load it into memory and start the application.

The signature validation and the public keys needed for the validation are all managed by the \ac{TEE}. The public keys or certificates are written into the \ac{TEE} in the \ac{IoT} device manufacturing process.
There may also be the possibility to securely update a public key or revoke a key.
The \ac{IoT} device manufacturer retains the private key that is used to digitally sign the firmware.
As long as the private signing key remains secure, this process should guarantee that only the device manufacturer can update the firmware and therefore the firmware can be trusted.
However some issues still remain, for example if the device manufacturer has not removed the debug interface, it may be possible to use this interface to change or add new keys~\cite{shwartz_reverse_2018}.

The secure boot process is part of the root of trust~\cite{zhao_providing_2014}, proving that the software running on the device is as expected and the keys being used are certified by the \ac{PKI}.
Without a secure boot process, the software on the \ac{IoT} device could have been modified, contain malware, or have modified keys. 

\subsection{Device identity key}
In addition to each device having a unique identification number or string, a device should be able to authenticate its identity as part of an attestation process.
This requires the \ac{IoT} device to have a private key that it can use to digitally sign a message.
Any entity communicating with the \ac{IoT} device can use the public key of that \ac{IoT} device to validate the digital signature and so can confirm the identity of the \ac{IoT} device as the \ac{IoT} device must be in possession of the private key. 
This private key is known as the device identity key.
The key can be written into the \ac{IoT} device when it is manufactured or at a later stage, known as device provisioning. 

A public-private key pair is generated, with the private key being stored within the \ac{TEE}, and the public key shared with multiple third parties.
A certificate for the public key can also be generated and signed by a trusted \ac{CA}.
The private key is then used to generate a signature to attest ownership of the identity, for example using the \ac{PSA} Attestation \ac{API}~\cite{arm_psa_2022}.
The process of generating the keys may be done as part of the manufacturing process, as part of device provisioning, or when a device enlists in a service.
Platform manufacturers that provide an \ac{IoT} cloud platform may have a provisioning process, in which a public and private keys are generated (possibly on the device) and then the private key is stored in the \ac{TEE}.
The public key certificate is then stored in the cloud service as part of the device description and is later used in the authentication process. 
Other platform manufacturers provide tools that allow the \ac{IoT} device manufacturer to prevision a device themselves.
The platform manufacturers may also ship with a private key embedded in microprocessor or \ac{TEE}, with the public key being included as part of the documentation.
The \ac{IoT} device manufacturer can then register the public key for the device with cloud services.

\subsection{Unique per device password}
An \ac{IoT} device may require the user to log in to an administration interface to configure certain features.
This administration account should not use a default or well known device password, but should be secured with a unique per device password that is affixed to the device.
Users should be able to change this default password. This new password should be stored securely, adhering to standards such as NIST SP 800-63 standard~\cite{grassi_digital_2020}. The majority of users may not change their default password~\cite{bosnjak_what_2016}, making it more important that any unique per device password follow good password practice, for example NIST SP 800-63B~\cite{grassi_digital_2017}, or three random words~\cite{ncsc_logic_2021}. 
Factory resetting the device should also reset the password back to the unique per device password.
The unique per device password should be generated at manufacturing time and stored in the \ac{IoT} device, possibly in the \ac{TEE} depending on the hardware functionality.
The unique per device password is not a cryptographic key, it should be something that a users should be able to remember, but that an attacker should find hard to guess.
The unique per device password should be generated randomly and not be influenced by known information such as the device ID or MAC address, or follow a pattern for each device. 
\section{Method}
To generate a list of manufacturers and \ac{IoT} platforms, the initial list of manufacturers was derived from the authors' knowledge of the \ac{IoT} industry.
For each platform in the initial list, an internet search was made to find competitors or similar \ac{IoT} platforms from other manufacturers.
We focused on devices that may be used within the consumer market, these devices tend to be initially configured by a user connecting with a phone via Bluetooth to the \ac{IoT} device.
Once the initial configuration is done, the \ac{IoT} device will then connect to the Wi-Fi network.
Therefore, we selected devices that have both Wi-Fi and Bluetooth functionality. 
We also limited the selection to \ac{IoT} platforms that use a \ac{SoC} 32bit microprocessor.
We only included devices where the platform manufacturer has provided a development or prototyping board.
These prototyping boards give \ac{IoT} device engineers the ability to explore the functionality of the \ac{IoT} platform and build the application without needing to design and build the \ac{IoT} device \ac{PCB}.
The prototyping boards come with debug and programming interfaces allowing them to be connected to a PC. This make the platform more accessible and allows the platform manufacturer to target a wider user base. 
For the review we limited the platform selection to the following criteria:
\begin{itemize}
  \raggedright
  \item Platform manufacturers targeted a specific product for the \ac{IoT} device market.
  \item The platform was a \ac{SoC} with a 32bit microprocessor.
  \item The platform included both Wi-Fi and Bluetooth connectivity.
  \item A development or prototyping board was available to evaluate the functionality of the \ac{SoC} \ac{IoT} platform.
  \item The documentation was available without needing to be a member of a trade organisation or signing any form of \ac{NDA}.
\end{itemize}
The list of \ac{IoT} platforms reviewed is shown in Table~\ref{tab:table_1}. For each platform we reviewed the following:
\begin{itemize}
  \raggedright
  \item The prototyping board data-sheet, the \ac{SoC} data-sheet and any additional or peripheral data-sheets.
  \item The \ac{IDE}.
  \item Programming guides for the following:
    \begin{itemize}
      \item General security coding guidance or good practice.
      \item Secure boot, root of trust and cryptographical operations.
      \item Device identity keys and key management.
      \item Generating unique per device passwords, password management or storage.
      \item Guidance on storage of sensitive information such as passwords.
    \end{itemize}
  \item Code examples involving any aspect of the three security features.
\end{itemize}
\begin{table*}
  \centering
  \begin{tabular}{|lllllll|} 
    \hline
    Manufacturer&  Product&  Chipset&  CPU&  Secure element&  \begin{tabular}{@{}c@{}}Secure\\ Boot\end{tabular}& \begin{tabular}{@{}c@{}}Device\\ Identity\\ Key\end{tabular}\\ 
    \hline
    Arduino&  Portenta&  STM32H747XI&  \begin{tabular}{@{}l@{}}Arm Cortex-M7\\ Arm Cortex-M4\end{tabular}&  \begin{tabular}{@{}l@{}}Microchip ATECC608\\ NXP EdgeLock SE050C2\end{tabular}&  Yes&  Yes\\ 
    \hline
    Espressif&  ESP32&  ESP32&  Tensilica Xtensa&  &  Yes&  Yes\\ 
    \hline
    Infineon&  PSoC™ 64&  CY8C62xA&  \begin{tabular}{@{}l@{}}Arm Cortex-M4\\ Arm Cortex-M0+\end{tabular}&  \begin{tabular}{@{}l@{}}Arm Cortex-M10\\Cryptography\\ Accelerator\end{tabular}&  Yes&  Yes\\ 
    \hline
    Nordic&  nRF7002 DK&  nRF54&  \begin{tabular}{@{}l@{}}Arm Cortex-M\\ Arm Cortex-M33\end{tabular}&  Arm TrustZone&  Yes&  Yes\\ 
    \hline
    Particle&  Photon 2&  \begin{tabular}{@{}l@{}}Realtek7\\ RTL8721DM\end{tabular}&  Arm Cortex-M33&  Arm TrustZone&  Yes&  Yes\\ 
    \hline
    Raspberry Pi&  Zero 2 W&  \begin{tabular}{@{}l@{}}Broadcom\\ BCM2710A\end{tabular}&  Arm Cortex-A53&  None&  No&  No\\ 
    \hline
    Pine64&  PineNut&  \begin{tabular}{@{}l@{}}bouffalolab\\ BL602 SoC\end{tabular}&  &  &  Yes&  No\\ 
    \hline
    Microchip&  WFI32-IoT&  PIC32&  PIC32MZ W1 MCU&  Trust\&Go&  No&  Yes\\ 
    \hline
    Silicon labs&  SiWx917&  &  ARM Cortex-M4&  ThreadArch&  Yes&  Yes\\ 
    \hline
    \multicolumn{7}{l}{Yes - The security feature was supported by the platform}\\
    \multicolumn{7}{l}{No - The security feature was not supported by the platform}\\
    \end{tabular}
  \caption{\ac{IoT} platforms reviewed}
  \label{tab:table_1}
\end{table*}

\noindent
Having completed the initial review, the evaluation criteria were constructed, and used to evaluate each set of material.

\subsection{Evaluation criteria}
The review includes documentation, code examples, libraries and tools provided by the platform manufacturers.
Where the platform manufacturers have included direct links to the documentation provided by the manufacturer of a microprocessor or secure element used in the \ac{IoT} platform, we have included them in the documentation set for the \ac{IoT} platform manufacturer.
We have not included code examples from other third parties, blog posts, or post community sites (including community sites owned by the platform manufacturers).
We placed each review item into the following categories based on the criteria below:

\subsubsection{Supporting Documentation}
The documentation provides an introduction to the security feature, providing the reader with an understanding of the security feature and why it is important for the \ac{IoT} developer to include the security feature in their implementation.

\subsubsection{Standards or regulations}
The documentation refers to the relevant recommendation, standard, regulation or certification for the security feature.
The documentation explains why the standard is relevant, how the standard should be applied and any additional consideration or actions that the device engineer should perform.

\subsubsection{Technical detail}
The documentation provides a detailed technical description of the security feature. The documentation includes implementation detail and explains the internal operation of the security feature. Technical detail include data-sheets describing how the hardware supports the security feature.

\subsubsection{Developer support}
The documentation provides detail of how the security feature can be implemented using the \ac{IoT} platform hardware or software, including \ac{API} documentation, examples of code, and configuration settings.
The documentation has a step-by-step guide to using the platform to achieve the security feature.

\subsubsection{Code examples}
The platform manufacturer has provided example source code of how the security feature can be implemented using the hardware.
This should also include instructions for any device configuration, building and running the example code. 
This example source code should also follow security best practice.

\subsubsection{Library}
The platform manufacturer provides a library that supports the implementation of the security feature.
This library may include an \ac{API} that the device engineer can use to gain access to the security feature.

\subsubsection{Ancillary code}
The platform manufacturer provides additional libraries, code or tools that can be used to support the implementation of the security feature. For example, additional key management, generation of strong passwords, or certification management.

\subsection{Threats to validity}
There are many manufacturers providing platforms for the \ac{IoT} market, often with multiple variations of each product, some targeting niche markets or very specific requirements.
Manufacturers may also provide a more componentised approach where different peripherals are added to a base computing platform, for example adding a Wi-Fi co-processor to an existing product to add Wi-Fi connectivity.
\ac{IoT} device manufacturers may choose to take this componentised approach if they do not plan to use all the features of the \ac{SoC} or have other constrains such as power consumption, cost \textit{etc}.
Given the size and complexity of this market and the number of different variations, this review has only taken a thin slice of the possible set of all \ac{IoT} device components.
We have used a selection criterion that only includes \ac{SoC} with both Wi-Fi and Bluetooth, and where the manufacturer provides prototyping or development boards.
These criteria limited the number of devices in the review and may skew the findings as we may have missed some interesting examples.

Acar {et al.} (2019)~\cite{acar_you_2016} have shown that developers often use question-and-answer sites to find solutions for security questions; the study showed that developers using the documentation have more secure implementations but take longer writing the feature.
Following from this work, we focused on the platform manufacturers' documentation and example code and did not review corresponding developer community sites. Platform manufacturers such as Arduino or Raspberry Pi focus more on the educational or hobbyist market and may also rely more on the user community creating content. For this paper we have not reviewed the community content.

We focused the review on three example security recommendations from the standards and regulations.
We did not look at issues such as data security and privacy or other hardware features such as flash encryption; this may limit the generalisation of this work. 

\section{Results}
Table \ref{tab:table_2} contains the evaluation of \ac{IoT} platforms.
The platform manufacturers take different approaches to helping \ac{IoT} device manufacturers and their engineers; we categorise the approaches as:
\begin{itemize}
  \item Cloud services deployment
  \item \ac{OS} and security features support
  \item Data-sheets
  \item Limited security features
\end{itemize}

\begin{table*}
  \centering
  \begin{tabular}{|r*{21}{|p{3.7mm}}|}
    \hline
    & \multicolumn{3}{|c|}{\begin{tabular}{@{}c@{}}Supporting\\ Documentation\end{tabular}
    }&  \multicolumn{3}{|c|}{\begin{tabular}{@{}c@{}}Standards or\\ regulations\end{tabular}}&  \multicolumn{3}{|c|}{\begin{tabular}{@{}c@{}}Technical\\ detail\end{tabular}}&  \multicolumn{3}{|c|}{\begin{tabular}{@{}c@{}}Developer\\ support\end{tabular}}&  \multicolumn{3}{|c|}{\begin{tabular}{@{}c@{}}Code\\ examples\end{tabular}}&  \multicolumn{3}{|c|}{Library}&  \multicolumn{3}{|c|}{\begin{tabular}{@{}c@{}}Ancillary\\ code\end{tabular}}\\
    Manufacture&  sb&  id&  pw&  sb&  id&  pw&  sb&  id&  pw&  sb&  id&  pw&  sb&  id&  pw&  sb&  id&  pw&  sb&  id&  pw\\
    \hline
    Arduino&      \Circle& \Circle& \Circle&   \CIRCLE& \CIRCLE& \Circle&   \CIRCLE& \Circle& \Circle&   \CIRCLE& \LEFTcircle& \Circle&  \Circle& \LEFTcircle& \Circle&   \CIRCLE& \Circle& \Circle& \Circle& \Circle& \Circle\\
    Espressif&    \CIRCLE& \Circle& \Circle&   \Circle& \Circle& \Circle&   \CIRCLE& \CIRCLE& \Circle&   \Circle& \CIRCLE& \Circle&    \Circle& \CIRCLE& \Circle&    \Circle& \Circle& \Circle&   \CIRCLE& \Circle& \Circle\\
    Infineon&     \Circle& \Circle& \Circle&   \CIRCLE& \CIRCLE& \Circle&   \CIRCLE& \CIRCLE& \Circle&   \CIRCLE& \CIRCLE& \Circle&    \CIRCLE& \CIRCLE& \Circle&    \Circle& \Circle& \Circle&   \CIRCLE &\CIRCLE& \Circle\\
    Nordic&       \CIRCLE& \CIRCLE& \Circle&   \CIRCLE& \CIRCLE& \Circle&    \CIRCLE& \CIRCLE& \Circle&   \CIRCLE& \CIRCLE& \Circle&   \CIRCLE& \CIRCLE& \Circle&     \CIRCLE& \CIRCLE& \Circle&   \Circle& \Circle& \Circle\\
    Particle&     \CIRCLE& \LEFTcircle& \Circle&   \Circle& \Circle& \Circle&   \CIRCLE& \CIRCLE& \Circle&    \Circle& \Circle& \Circle&    \Circle& \Circle& \Circle&   \Circle& \Circle& \Circle& \Circle& \Circle& \Circle\\
    Raspberry pi& \Circle& \Circle& \Circle&   \Circle& \Circle& \Circle&    \CIRCLE& \Circle& \Circle&   \Circle& \Circle& \Circle&    \Circle& \Circle& \Circle&    \Circle& \Circle& \Circle&   \Circle& \Circle& \Circle\\
    Pine64&       \Circle& \Circle& \Circle&   \Circle& \Circle& \Circle&    \CIRCLE& \Circle& \Circle&   \Circle& \Circle& \Circle&    \Circle& \Circle& \Circle&    \Circle& \Circle& \Circle&   \Circle& \Circle& \Circle\\
    Microchip&    \Circle& \CIRCLE& \Circle&   \Circle& \Circle& \Circle&    \Circle& \CIRCLE& \Circle&   \Circle& \LEFTcircle& \Circle&  \Circle& \Circle& \Circle&   \Circle& \Circle& \Circle&   \Circle& \Circle& \Circle\\
    Silicon labs& \Circle& \Circle& \Circle&   \Circle& \Circle& \Circle&    \CIRCLE& \Circle& \Circle&   \CIRCLE& \Circle& \Circle&     \Circle& \Circle& \Circle&    \Circle& \Circle& \Circle&   \Circle& \Circle& \Circle\\
    \hline
    \multicolumn{22}{l}{\CIRCLE - Material discussed the security feature}\\
    \multicolumn{22}{l}{\LEFTcircle - Material discussed the security feature as part of a cloud services}\\
    \multicolumn{22}{l}{\Circle - Material did not discuss the security feature}\\
    \multicolumn{22}{l}{sb - Secure Boot, id - Device identity key, pw - Unique per device password}\\
  \end{tabular}
  \caption{Evaluation of \ac{IoT} platform documentation}
  \label{tab:table_2}
\end{table*}

\noindent
These approaches are discussed below.
\subsection{Development support approaches}
\subsubsection{Cloud services deployment}
\begin{figure*}
  \includegraphics[width=\textwidth]{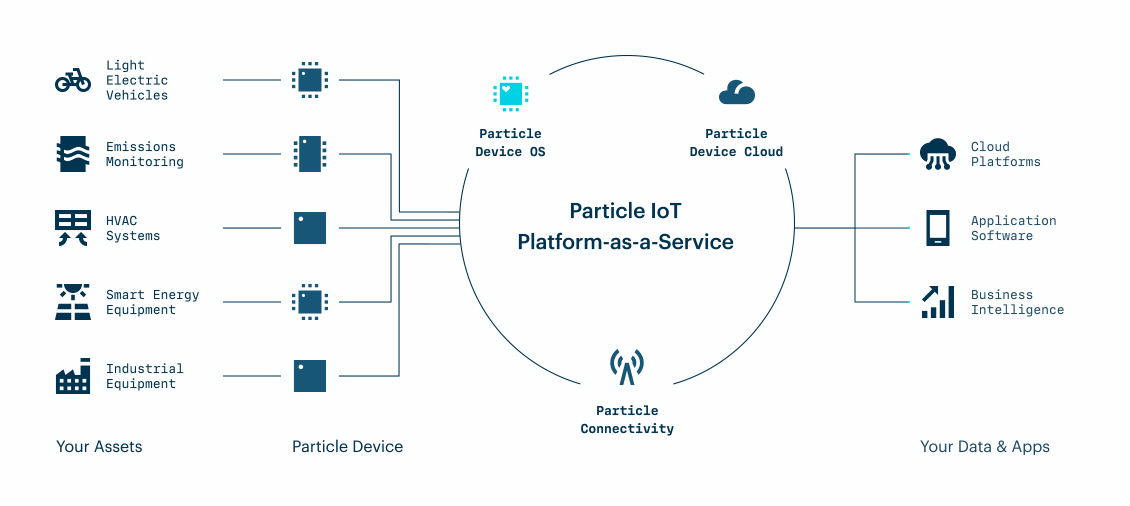}
  \Description[Particle PaS]{The Particle cloud service with multiple different \ac{IoT} device connecting to cloud services.}
  \caption{Particle Platform-as-a-Service~\cite{particle_device_2024}}
  \label{fig:pas}
\end{figure*}

In this category the platform manufacturers attempt to remove some of the security burden by providing a platform that includes: the \ac{SoC} hardware, a device \ac{OS}, security features and cloud infrastructure that manages the devices and provides features for the \ac{IoT} device engineers. For example the Particle Platform-as-a-Service in Figure~\ref{fig:pas}.
The platform manufacturers provide a root of trust via the cloud infrastructure, using a provisioning process to generate and securely store keys.
The \ac{IoT} device engineers use the integrated development environment to build and sign the application firmware. The focus from the platform manufacturers is on their secure development process and providing a secure platform that device engineers can simply use.
Third party software vendors are also providing \ac{IoT} cloud services for example, AWS \ac{IoT} Core~\cite{aws_secure_2024}. These services aim to allow any \ac{IoT} device to use the services, with some platform manufacturers providing examples of how to integrate their device with the third party service.
For some \ac{IoT} device engineers there are clear advantages for this approach and it can be useful for prototype development and small volume production.
Platform manufacturers such as Arduino and Particle have offerings in this space, but we can see other platform manufacturers moving in this direction.
Using these services introduces supply chain issues: consider, an \ac{IoT} device collecting personal sensitive information. This information may be processed on the \ac{IoT} cloud platform and stored using a cloud service provider, both actions may take place in different geo-political locations.
This can have both security and data protection ramifications that \ac{IoT} device manufacturers will need to consider~\cite{abba_ari_enabling_2020}.
The cost to a \ac{IoT} device manufacturer could be significant for a large-scale deployment. \ac{IoT} device manufacturers may also need to consider issues such as  vendor lock-in or over reliance on a single cloud service provider.
\subsubsection{\ac{OS} and security features support}
\begin{figure*}
  \begin{subfigure}{0.49\textwidth}
    \includegraphics[width=1.0\linewidth]{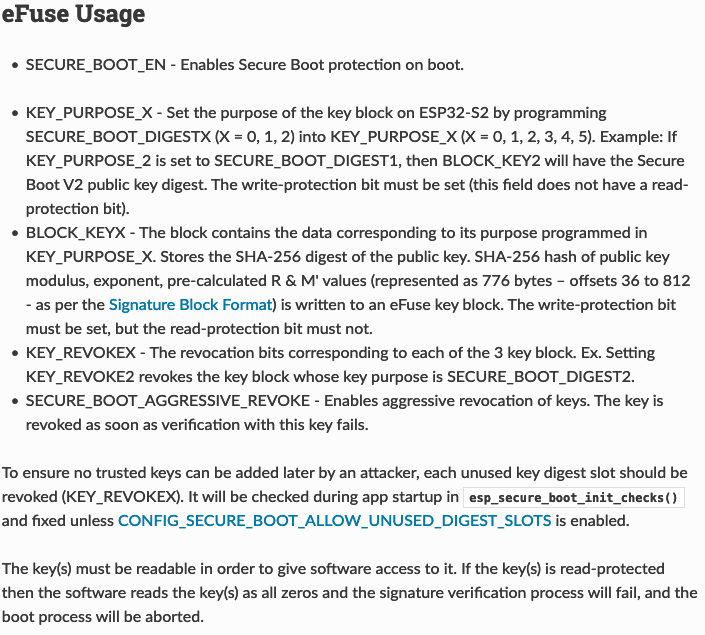}
    \caption{Espressif, storing Keys for secure boot~\cite{espressif_secure_2024}. }
    \label{fig:espressif_eFuse}
  \end{subfigure}
  \begin{subfigure}{0.49\textwidth}
    \includegraphics[width=1.0\linewidth]{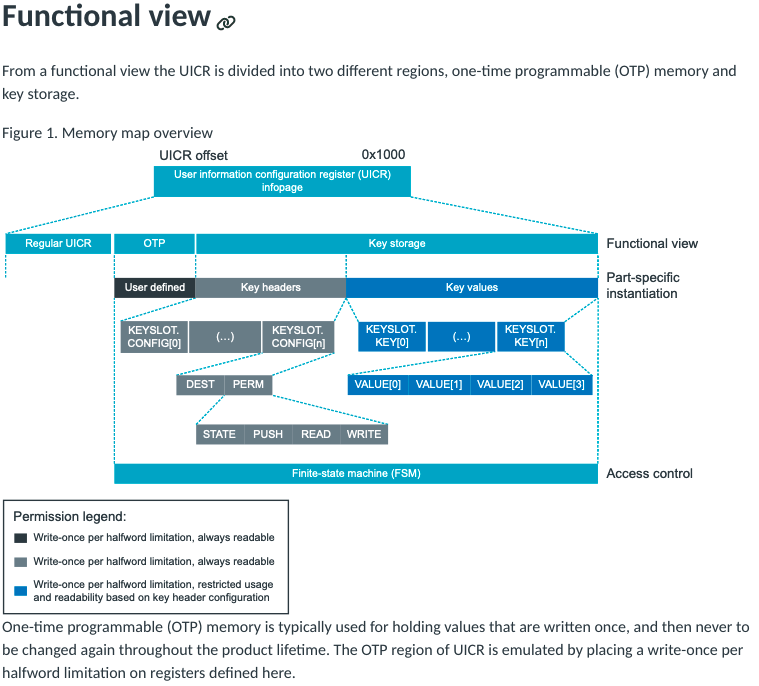}
    \caption{Nordic, key management unit~\cite{nordic_kmu_2024}}
    \label{fig:Nordic_key_managemen}
  \end{subfigure}
  \begin{subfigure}{1.0\textwidth}
    \includegraphics[width=1.0\linewidth]{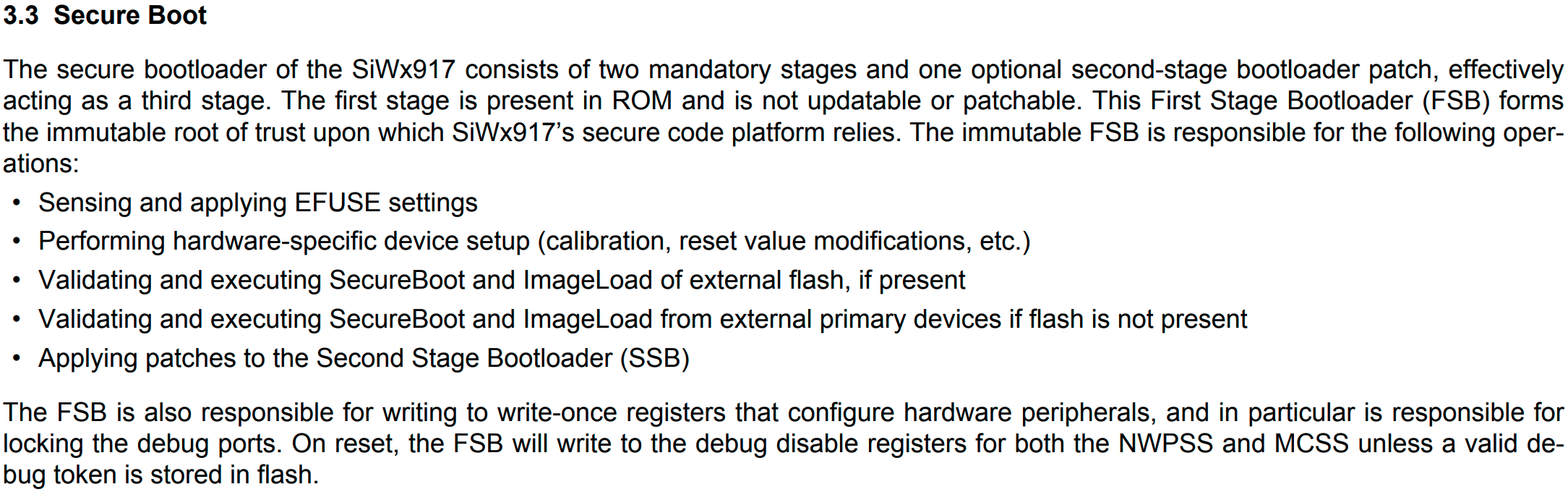}
    \caption{Silicon labs, secure boot~\cite{silicon_labs_an1428_2024}.}
    \label{fig:Silicon_labs_Secure_boot}
  \end{subfigure}
  \Description[technical documentation]{Three examples of technical documentation of security functionality.}
  \caption{Example of security functionality technical documentation}
  \label{fig:data-sheets}
\end{figure*}

The platform provides the \ac{SoC}, a real time \ac{OS}, such as FreeRTOS~\cite{freertos_freertos_2024} or zephyr~\cite{zephyr_project_zephyr_2024} and a \ac{TEE}.
The security functionality is implemented within the \ac{TEE} using an environment such as TrustedFirmware-M~\cite{arm_trusted_2022}.
The aim is to provide an ecosystem that has all the parts that a \ac{IoT} device engineer requires to build secure \ac{IoT} devices.
This strategy relies on good documentation for the device engineers and access to example code that demonstrates the correct use of the security functionality.
All of the platform manufacturers have some technical detail normally delivered as data-sheets describing at least one of the security features.
The level of detail contained within the technical documentation can very significantly.
This documentation is intended to describe the functionality of the \ac{IoT} platform. Additional material such as developer support and example code is needed to support a device engineer using the hardware functionality to deliver a security feature.  
Figure~\ref{fig:data-sheets} has some examples taken from technical documentation. 
These example show different levels of detail that can be found within the technical documentation, Figure~\ref{fig:espressif_eFuse} has an example of an overview of the security functionality. More information or example code may be needed for device engineers to take advantage of these functionality. In contrast Figure~\ref{fig:Nordic_key_managemen} is the beginning of a detailed explanation of the key management unit and how a device engineer can use the security functionality. 

\begin{figure*}
  \begin{subfigure}{0.49\textwidth}
    \includegraphics[width=0.9\linewidth]{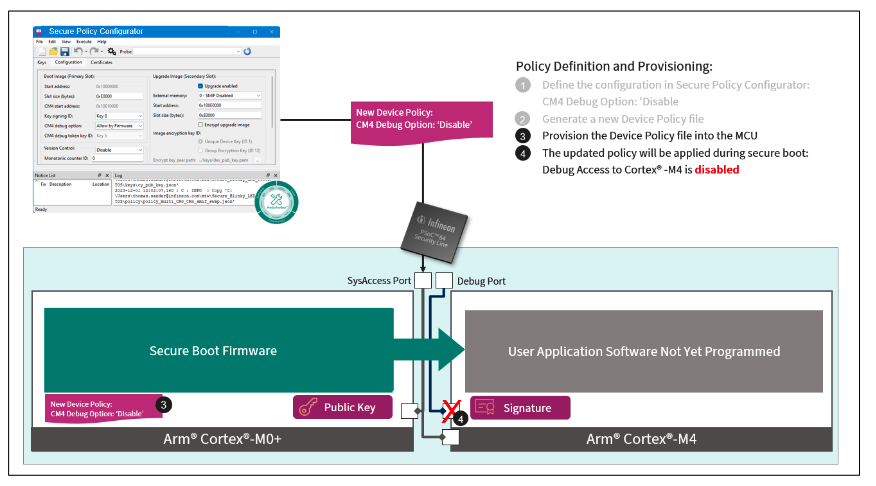}
    \caption{Infineon, secure policy configurator~\cite{infineon_psoc_2023}. }
    \label{fig:infineon_psoc}
  \end{subfigure}
  \begin{subfigure}{0.49\textwidth}
    \includegraphics[width=0.9\linewidth]{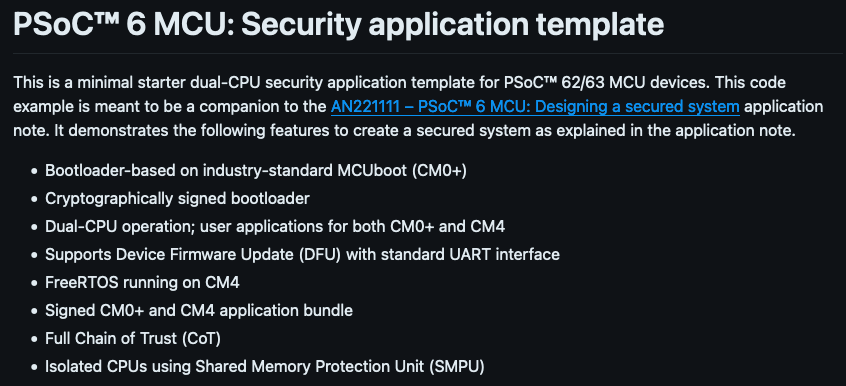}
    \caption{Infineon, Security application template~\cite{infineon_infineonmtb-example-psoc6-security_2024}.}
    \label{fig:Infineon_app_template}
  \end{subfigure}
  \Description[Developer support]{example of developer documentation and code example.}
  \caption{Example of developer documentation and code example.}
  \label{fig:dev-support}
\end{figure*}

Four of the platform manufacturers provided developer support documentation and two had code examples for the secure boot process. Figure~\ref{fig:infineon_psoc} is an example of the documentation explaining how the hardware is used to achieve a security feature, while Figure~\ref{fig:Infineon_app_template} is the start of the documentation for example code containing a security application template. 
The results were similar for device identity keys: five platform manufacturers providing developer support documentation and four having code examples.
We found no developer support documentation or code examples for unique per device passwords.
We also found that some of the code examples used hard-coded credentials or defaulted to a non-secure option.

\subsubsection{Data-sheets}
Some platform manufacturers provide the hardware for secure development but leave it to the \ac{IoT} device engineers to implement the real time \ac{OS} and other features.
This provides the device engineers with the flexibility to implement what is needed for their specific requirements.
The device engineers are reliant on the technical documentation provided by the platform manufacturers, this is usually in a data-sheet.
These data-sheets do not normally provide developer guides or example code. Eight of the platform manufacturers had data-sheets that described the operation and configuration of the secure boot process. Five of them had data-sheets describing device identity keys, with nothing covering unique per device passwords.

\subsubsection{Limited security features}
Platform manufacturers such as Raspberry Pi may utilise existing development platforms for the \ac{IoT} market. These existing platforms may not contain a \ac{TEE} and so cannot directly support some of the security requirements for this review. 

\subsection{Secure boot}
The secure boot process is supported by seven out of the nine platform manufacturers.
The documentation varied widely from one manufacturer having almost no detail, to three providing developer support documentation and example code. One of these also supplied a port of a code library to support the secure boot process. Three provided an introduction to the secure boot process and why it was important. Three  manufacturers introduced a standard, in all cases this was \ac{PSA} Certified~\cite{psacertified_psa_2023} as the secure boot process is part of the level one platform certification. Platform manufacturers can use \ac{PSA} Certified to demonstrate that the hardware meets an industry security certification. We did not see any reference to regulations such as the \ac{EU} Cyber Resilience Act (EU-CRA)~\cite{eu_cyber_2023}.

\subsection{Device identity key}
Device identity keys are supported by seven of the platform manufacturers; again we see a wide difference in the documentation for this security feature and how this security feature is presented to device engineers.

\subsubsection{Cloud services}
Platform manufacturers that supply cloud services demonstrate the use of a device identity key as part of the cloud provisioning process.
For example, Particle implements the integration with the \ac{IoT} device and the Particle cloud platform~\cite{particle_introduction_2024}.
Arduino delivers a simple sketch that is used to provision the \ac{IoT} device with the Arduino cloud platform \cite{pennasilico_arduinoiotcloudexamplesutilityprovisioning_2024}.
Microchip generates a key using a \ac{PUF} as part of the manufacturing process. They include multiple examples of the use of device identity keys with third party \ac{IoT} cloud services such as AWS \ac{IoT} Core~\cite{microschip_securely_2023}  or Google cloud~\cite{passemard_securing_2018}.

\subsubsection{Data sheets}
\begin{figure}
  \begin{subfigure}{\linewidth}
    \includegraphics[width=\linewidth]{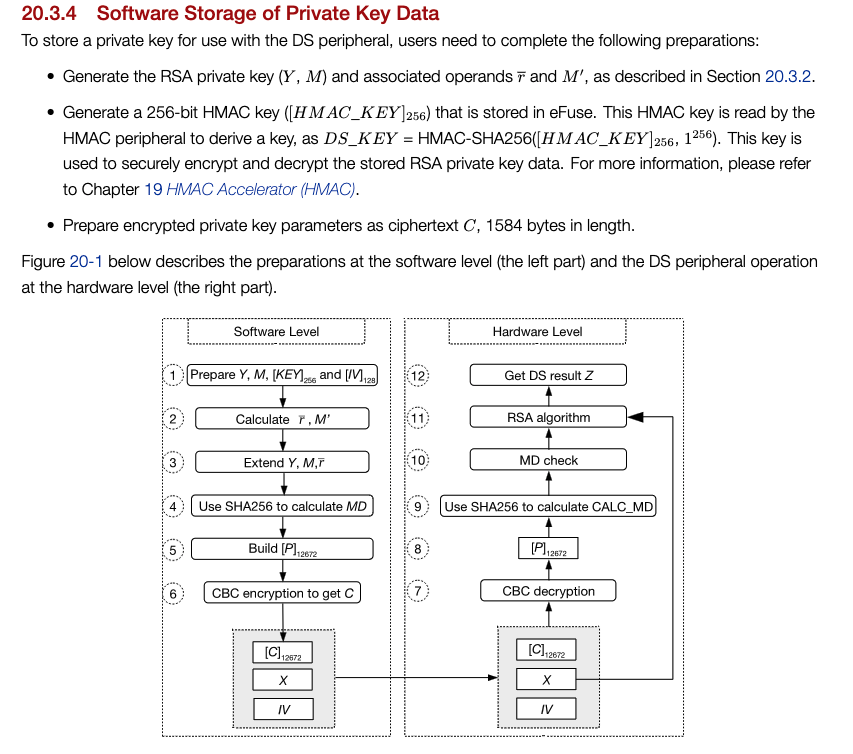}
    \caption{Espressif, digital signature operation~\cite{espressif_esp32-s2_technical_reference_manual_enpdf_2023}. }
    \label{fig:subim6}
  \end{subfigure}
  \begin{subfigure}{\linewidth}
    \includegraphics[width=\linewidth]{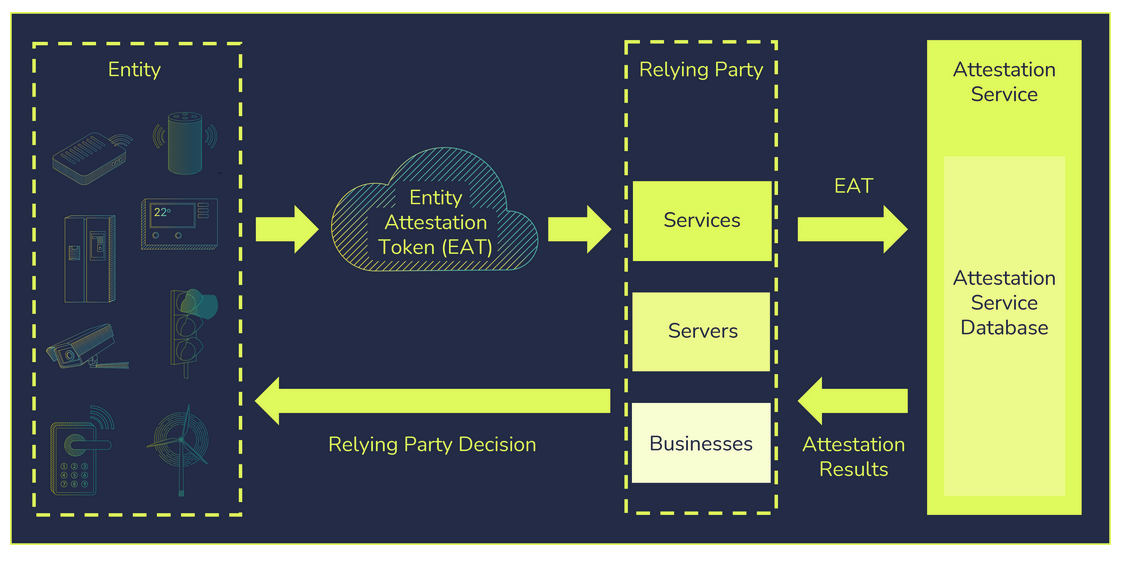}
    \caption{PAS Certified, device attestation~\cite{psacertified_attestation_2021}.}
    \label{fig:subim7}
  \end{subfigure}
  \Description[Security functionality]{example of data-sheet with technical explanation of a security functionality and the Attestation process that would use the functionality.}
  \caption{Example of security functionality and the security feature to be implemented using the functionality.}
  \label{fig:attestation}
\end{figure}

Five platform manufacturers supply technical details of secure boot or device identity key, without providing additional developer support.
This technical documentation focuses on the hardware functionality, not how to implement services like provisioning or attestation. See Figure~\ref{fig:attestation} for an example of a data-sheet explaining the use of digital signatures using the private key to sign a message.
Figure~\ref{fig:attestation} also has the overview of the device attestation process;
the device engineer must bridge the gap from the functional definition of the hardware and software support provided by the \ac{IoT} platform and the security feature to be implemented, in this case device attestation using a device identity key.

\subsection{Unique per device password}
Some of the platform manufacturers did provide more general security-related development guidelines or certification programs such as \ac{PSA} Certified.
These certification programs provide a framework and checklist for the \ac{IoT} device engineers.
If the \ac{IoT} platform is also certified, that can then support the certification of the \ac{IoT} device.
Implementing the unique per device password recommendations necessitates a number of requirements:
\begin{itemize}
  \item A mechanism to generate a unique per device password, at manufacturing time, that can be stored in the \ac{IoT} device firmware and affixed to the outside of the device.
  \item A mechanism to securely store the unique per device password. The password stored in a protected or tamper-proof zone, possibly in the \ac{TEE}. 
  \item A mechanism to override the device password with a user defined password. This also requires that user password is stored securely and should meet good password recommendations such as NIST SP 800-63B~\cite{grassi_digital_2017}.
  \item A mechanism to reset the device password back to the original unique per device password as part of the factory reset process.
\end{itemize}
Implementing these requirements entails understanding the \ac{IoT} platform security functionality, the recommendations or regulations covering the \ac{IoT} device and how the customer will use the device.
\ac{IoT} device engineers may wish to take advantage of the \ac{IoT} platform security functionality, such as the storage of sensitive information or the use of cryptographic functions to impalement some of the features above. 
We did not find any evidence from the platform documentation for recommendations or support for providing unique per device passwords.
Some platform manufacturers do supply tools to run at manufacturing time to write keys or other data, these could be used as a template to write a unique per device password at manufacturing time.
Some of the platforms reviewed do present hardware support for storing sensitive data, but we found no discussion around how this could be used other than for key storage.

Naiakshina \textit{et al.} (2019)~\cite{naiakshina_if_2019} asked 42 freelance developers to store a password, 17 (40\%) provide a secure solution.
Acar \textit{et al.} (2017)~\cite{acar_comparing_2017} looked at the use of cryptographical library \ac{API}s. Having a simple interface was good, but they should also include ancillary functions to help the developers use the library.
We may conclude that if we want \ac{IoT} device engineers to securely store and manage passwords, we should add these ancillary functions to existing cryptographic libraries that the \ac{TEE} devices already provide. 

\subsection{Device engineer support}
Eight of the nine platform manufacturers are providing hardware support for some basic \ac{IoT} security tasks. Five platforms include a \ac{TEE}, with three others having mechanisms to support security functionality such as the secure boot process or device identity keys. Platform manufacturers are also providing cryptographic libraries or hardware acceleration, and other features such as encrypted firmware. The security features provided do vary across manufacturers. Certification programs such as \ac{PSA} Certified are driving some level of conformity but this standard may not be suitable for all \ac{IoT} platform manufacturers.

The technical detail, developer support and code examples provided by different platform manufacturers varies significantly across manufacturers. Five of the nine platform manufacturers provide developer support or code examples.
Platform manufacturers focus on the hardware that they deliver and the security functionality of that hardware. 
Platform manufacturers do not address the needs of device engineers that must implement regulatory requirements  such as unique per device password.

\subsubsection{Additional support provided}
Platform manufacturers take different approaches to providing additional support for device engineers to take advantage of the security functionality that they deliver.
Three platform manufacturers provide cloud solutions that attempt to shield the device engineers from the implementation details of the security features.
These solutions offer features like secure boot, device identity, and provisioning within the operations of the cloud service. Two platform manufacturers provide additional functionality using libraries or \ac{API}s that makes use of the hardware. For example the TrustedFirmware-M~\cite{arm_trusted_2022} Initial Attestation Service Integration Guide is implemented on the Arm TrustZone~\cite{arm_trustzone_nodate}. The \ac{API} delivers a set of security features that can be used by device engineers without the need to deal directly with the low level hardware.  Some manufacturers take a componentised approach, providing discrete hardware components, that an \ac{IoT} device manufacture would select to best meet their individual market or technical needs.

This additional support does not extend past the security primitives, to provide support for regulatory recruitment or \ac{IoT} standards recommendations. 

\section{Related work}
\ac{IoT} security has been a research focus area for some time; Alqassem \& Svetinovic (2014)~\cite{alqassem_taxonomy_2014} proposed a taxonomy for \ac{IoT} security and privacy requirements.
As part of a systematic review, Mohanty \textit{et al.} (2021)~\cite{panigrahi_iot_2021} categorised the \ac{IoT} security challenges, focusing on \ac{IoT} architecture and protocols.
Mishra \& Pandya (2021)~\cite{mishra_internet_2021} reviewed the security challenges and solutions including \ac{IDS}, while Mohanta \textit{et al.} (2020)  considered security challenges and solutions using machine learning and block chain technology.
Much of the research looks at the technical security challenges for \ac{IoT} devices; Schiller \textit{et al.} (2022)~\cite{schiller_landscape_2022} take a wide view of the overall \ac{IoT} security landscape, concluding that \ac{IoT} security remains a concern given issues like limited resources or the need for fast time to market, though there are now devices on the market that can make the use of \ac{IoT} devices more secure.
Pinto \& Santos (2019)~\cite{pinto_demystifying_2019} reviewed the introduction of the Arm TrustZone~\cite{arm_trustzone_nodate} as a \ac{TEE} within the \ac{IoT} market. Ling \textit{et al.} (2021)~\cite{ling_secure_2021} looked at the use of TrustZone for \ac{IoT} devices and demonstrated how it can be used for trusted boot and remote attestation. 

Chowdhury \textit{et al.} (2021)~\cite{chowdhury_developers_2021} considered the issues being faced by developers when writing security-related functionality; they defined a number of developer challenges, behaviors, and interventions as well as a set of tropes (something that is considered true, but is not).
These tropes include the notion that the developer is an expert.
They may be an expert in software engineering, but that does not necessarily make them an expert in security, cryptography or privacy.
Yskout \textit{et al.} (2015)~\cite{yskout_security_2015} showed that development teams did not seem to perform any better when following security design patterns and that both teams suffered from a lack of detail, for example needing to manage keys when writing encrypted storage features, suggesting the need for supporting features or ancillary code.
When using privacy by design Senarath \& Arachchilage (2018)~\cite{senarath_why_2018} noted that developers struggled to map privacy requirements to engineering practice.
Naiakshina \textit{et al.} (2019)~\cite{naiakshina_if_2019} asked freelance developers to store a password; even when prompted to do this securely, 38\% failed to provide a secure solution.
Hallett \textit{et al.} (2021) went on to show that when the participants were asked to write a specification for the function to securely store a password before writing the code, only 3\% (two participants) followed the current best practice guidelines.
Acar \textit{et al.} (2017)~\cite{acar_comparing_2017} concluded that for cryptography libraries to reduce developer errors, it was not sufficient for the library to be simple to use.
The libraries also needed to include ancillary supporting functions, documentation and example code.
Acar \textit{et al.} (2016)~\cite{acar_you_2016} demonstrated that where developers get their security solutions from can affect the security of the resulting code.
Using sources such as StackOverflow can result in less secure solutions. 

\section{Discussion}
The regulatory and standards landscape for \ac{IoT} device engineers is complicated. 
Streamlining these standards is made difficult given the breadth of application areas that \ac{IoT} devices are deployed in, resulting in multiple standard bodies and differing requirements.
Organizations such as \ac{PSA} Certified~\cite{psacertified_psa_2023} provide a level of guidance to \ac{IoT} device manufacturers and a certification checklist to demonstrate that the \ac{IoT} device and development process meets a minimal standard.
Platform manufacturers are now including a \ac{TEE} within the \ac{IoT} device.
These \ac{TEE}s deliver a set of security primitives such as enabling a secure boot root of trust process, secure key and certificate management, and cryptographic functions.
Platform manufacturers are taking a number of different paths to address the needs of \ac{IoT} device manufacturers:
\begin{itemize}
  \item Delivering cloud-hosted services that integrate with the \ac{IoT} platform providing device management, provisioning and attestation.   
  \item Integration with third-party \ac{IoT} cloud services.
  \item Integrating with existing libraries for security functionality, such as Mcuboot~\cite{mcuboot_mcu-toolsmcuboot_2024} or TrustedFirmware-M~\cite{arm_trusted_2022}. 
  \item Providing development tools for signing images or writing keys or certificates into the \ac{IoT} device.
\end{itemize}
There remains a significant gap between the hardware security features provided by a \ac{TEE} and the functionality needed to follow the standards recommendations or to meet the requirements of the regulations.
Beyond the use of these security primitives, there is even less guidance for the \ac{IoT} device engineers to implement regulatory requirements such as password management.

\subsection{Standards landscape}
\begin{figure}
  \begin{framed}
    \begin{itemize} \raggedright 
      \item \ac{EU} General Data Protection Regulation (EU-GDPR)~\cite{eu_general_2016}.
      \item \ac{EU} Cybersecurity Act~\cite{eu_regulation_2019}.
      \item \ac{EU} Cyber Resilience Act (EU-CRA)~\cite{eu_cyber_2023}.
      \item \ac{UK} Product Security and Telecommunications Infrastructure (Product Security) regime~\cite{uk_uk_2024}.
      \item \ac{USA} \ac{IoT} Cybersecurity Improvement Act~\cite{rep_kelly_hr1668_2020}.
      \item California \ac{IoT} Security Law (SB-327)~\cite{california_bill_2020}.
    \end{itemize}
  \end{framed}
  \Description[IoT Regulation]{list of Regulations concerning IoT devices in the \ac{EU}, \ac{UK} and \ac{USA}}
 
  \caption{Regulations concerning \ac{IoT} devices in the \ac{EU}, \ac{UK} and \ac{USA}.}
  \label{fig:Standards_landscape}
\end{figure}
The regulatory landscape for \ac{IoT} device engineers is complex, with potentially multiple different regulations to consider, depending on the market, location of  deployment, and the data being collected.
Regulators are addressing the security concerns around \ac{IoT} devices by laying down a set of legal requirements for \ac{IoT} device engineers to follow (Figure~\ref{fig:Standards_landscape}).
It is the responsibility of the \ac{IoT} device engineer to convert these regulatory requirements into solutions that can be deployed in \ac{IoT} devices within a given set of constraints including cost, power, and size. 
Mapping the regulatory requirements to existing standards or recommendations is also challenging.
The ENISA Cyber Resilience Act Requirements Standards Mapping report~\cite{enisa_cyber_2024}, attempts to do this mapping for the \ac{EU} Cyber Resilience Act (EU-CRA)~\cite{eu_cyber_2023}.
They conclude that there is a need for a single unified set of \ac{IoT} recommendations, and that the mapping of regulatory requirements to existing standards either results in gaps or the need to refer to multiple standards. 
Looking at one example:
\begin{quote}
  \emph{``(3) On the basis of the risk assessment referred to in Article 10(2) and where applicable, products with digital elements shall:
  (a) be delivered with a secure by default configuration, including the possibility to reset the product to its original state;''}~\cite{eu_cyber_2023}
\end{quote}\leavevmode\newline
ENISA defines sub-requirements \cite{enisa_cyber_2024}:
\begin{itemize}
  \raggedright
  \item In case default configurations foresee an initial/default credential, the same should use a complex and randomly chosen password, different for each product
  \item In case default configurations cover cybersecurity items, they should adopt a reasonable level of security for each item
  \item The default configuration should be placed in a non-erasable memory
  \item A function to reset the product configuration to the default one should be implemented
\end{itemize}
The sub-requirements are mapped to other existing standards:
\begin{itemize}
  \raggedright
  \item ISO/IEC 27002 Information security, cybersecurity and privacy protection — Information security controls~\cite{1400-1700_isoiec_2022}
  \item ETSI EN 303 645 Cyber Security for Consumer Internet of Things: Baseline Requirements~\cite{etsi_cyber_2020}
  \item  ISO/IEC 18031:2011 Information technology — Security techniques — Random bit generation~\cite{iso_isoiec_2011}.
\end{itemize}

From the example above we can trace the requirement for a unique per device password, from the regulation via standard recommendations. Multiple standards would need to be consulted to implement this security feature.
Platform manufacturers provide no support for implementing this security feature.

The \ac{PSA}~\cite{arm_platform_2023} is an initiative lead by Arm to define a security architecture for \ac{IoT} devices.
The security model provides a set of industry best practices for \ac{IoT} platforms, defining a set of minimum security features that an \ac{IoT} device should conform to.
The \ac{PSA} Certified program~\cite{psacertified_psa_2023} enables \ac{IoT} device manufacturers to certify their product as following this standard.
Three of the platform manufacturers reviewed are part of the \ac{PSA} Certified program~\cite{psacertified_psa_2023}.
The \ac{PSA} Certified level 1 questionnaire~\cite{psacertified_psa_2023} also includes additional requirements to map the certification to the \ac{EU} Cyber Resiliency Act (EU-CRA)~\cite{eu_cyber_2023}.
Additionally, the level 1 requirements are also mapped to ETSI EN 303 645~\cite{etsi_cyber_2020}, this can be seen in Figure~\ref{fig:mapping}.
This again demonstrates the complexity of mapping regulations to standards and certifications.
This makes it difficult for both consumers and manufacturers to determine if a given regulation or standard is achieved by an \ac{IoT} device.
\begin{table*}
  \begin{tabularx}{\linewidth}{|X|X|}
    \hline
    \textbf{ETSI EN 303 645 V2.1.0 [2020-04) Provisions} &
    \textbf{PSA Level 1 Requirements} \\
    \hline
    5.1-1: Unique per device passwords & D4.1: Critical Security Parameters \\
    \hline
    5.1-2: Automated password attacks & D4.2: Automated password attacks \\
    \hline
    5.1-3: Cryptography for user authentication & S5.3 User authentication \\
    \hline
    5.1-4: Change of authentication value & S6.1: Security configuration \\
    \hline
    5.1-5: Authentication mechanism attack resilience & D4.2: Password best practices \\
    & D4.3: Password threshold \\
    \hline
    5.3-2: Mechanisms for secure updates & S1.1: Firmware update \\
    & S1.2: Anti-rollback \\
    \hline
    5.3-7: Best practice cryptography for updates & S1.1: Firmware update \\
    \hline
    5.3-10: Trust relationship for updates & S1.1: Firware update \\
    & D2.2: Client-Server Authentication \\
    \hline
    5.4-1: Sensitive parameter secure storage & S2.2: Secure storage \\
    \hline
    5.4-2: Secure storage of ID & C1.4: ID storage \\
    & D4.6: Storage \\
    \hline
    5.4-3: Configurable security parameters & D4.1: Critical security parameter \\
    \hline
    5.4-4: CSP unique per device restraint to automated attack & S5.1: CSP unique per device restraint to automated attack \\
    \hline
    5.5-1: Secure communication & S3.3: TLS \\
    \hline
    5.5-5: Authenticating parameter configuration & S6.1 Configuration \\
    & D3.10: Configuration \\
    \hline
    5.5-7: Sensitive data encryption over network & D2.3 Communication encryption \\
    \hline
    5.6-1: Disable unused ports & D2.1: No unused port \\
    \hline
    5.6-2: Minimize unauthorized disclosure & S3.3: Secure protocols that do not leak \\
    \hline
    5.6-4: Software disable of debug interface & S4.2: Unneeded functionalities\\
    \hline
  \end{tabularx}
  \caption{Mapping of ETSI EN 303 645~\cite{etsi_cyber_2020} requirements to \ac{PSA} Certified level 1 questionnaire~\cite{psacertified_psa_2023}}
  \label{fig:mapping}
\end{table*}

\subsection{Effect of regulation}
When reviewing the security functionality for these \ac{IoT} platforms, we see that the platform manufacturers are providing the security primitives: \ac{TEE}, root of trust, device identity keys, and cryptography functions. 
The regulations take a broad view, looking at the security needs for the life cycle of the \ac{IoT} device, including the development process, user data, vulnerability disclosure, and security updates, \textit{etc}.
These regulations define a set of processes that should be followed by the \ac{IoT} device manufacturing organisations, as well as security features that will need to be implemented on the \ac{IoT} device by device engineers.
Regulations may also place additional requirements on the \ac{IoT} hardware, and increase the need for more processing or storage.
For example, the need to log security events (such as a failed login attempt), or to interact with authorisation protocols. 

The implementation of a regulatory requirements may need the use of these security primitives. For example, encrypting data at rest or storing passwords in the \ac{TEE}.
We have seen from the example of unique per device passwords that minimal support is provided above the use of the security primitives to deliver these regulatory requirements.
There is a gap between the security provision from the platform manufacturers and the regulatory requirements.
Device engineers will need to implement these regulatory requirements themselves with the possibility that this will introduce other vulnerabilities.

\subsubsection{Cloud deployment}
The \ac{IoT} cloud deployments have the ability to remove much of the security detail from device engineers.
These platforms provide security services for root of trust, device identity, provisioning, and encrypted communication, as well as application features such as storage, message queues, and events.
We have seen from the review that the cloud services are delivering these security primitives.
There remains a gap between these security primitives and the regulatory requirements.
The cloud service providers have an opportunity to extend their offering to add additional regulatory functionality.

\subsection{Recommendations}
\subsubsection{Standards harmonisation}
Harmonising the current complexity of existing standards and recommendations would reduce confusion and misunderstanding.
Standard harmonisation could provide a single set of functional requirements that \ac{IoT} manufacturers could address, and encourage a unified certification program.
Harmonisation could also minimise the mapping from government regulations to multiple different \ac{IoT} standards.
Ultimately this may be beneficial to  platform manufacturers, \ac{IoT} device manufacturers, and to the consumer, potentially providing a single certificating mark that a consumer would recognize as being a certified secure device.
ETSI EN 303 645~\cite{etsi_cyber_2020} has been identified as a possible starting point for a single harmonised standard across the \ac{EU}~\cite{enisa_cyber_2024}.
The \ac{PSA} Certified program~\cite{psacertified_psa_2023} is attempting to take multiple standards and regulations into account. 
Continuing this work could make progress towards a harmonised standard, but work would be needed from multiple other regulators and standard bodies.

\subsubsection{Secure example code}
We find that developer support remains inconsistent across different platform manufacturers, with minimal step-by-step guides and code examples to implement a security feature.
We would encourage platform manufacturers to include more code examples.
These code examples should follow security best practices, and should not default to low or non-secure implementations. 
Platform manufacturers could also include code examples to implement specific standards recommendations.
For example, platform manufacturers that are \ac{PSA} Certified could also include code examples for \ac{PSA} Certified requirements, such as ID storage, password best practices, or security configuration.
This would aid device engineers in achieving \ac{PSA} certification for their \ac{IoT} device. It would also encourage the device engineers to follow the platform manufacturers' security coding practices and potentially provide a library of standards-based security functionality.
This would make it easier for device engineers to gain certification, which is also in the best interests of the platform manufacturers, as this will help to drive the selection of the \ac{IoT} platform.

\subsubsection{Software, and auxiliary code}
Platform manufacturers are already taking advantage of third-party or open source software such as FreeRTOS~\cite{freertos_freertos_2024} or TrustedFirmware-M~\cite{arm_trusted_2022} to provide functionality for \ac{IoT} devices.
This model could be extended to deliver functionality for regulatory requirements, for example by providing a library for password management and storage.
The \ac{IoT} platform manufacturer would be responsible for porting a set of libraries to their platform and keeping them up to date.
While this is additional work for the platform manufacturer, there is an  advantage; this would reduce the development cost of a new \ac{IoT} device, making the \ac{IoT} platform commercially more attractive, potentially  increasing thier market share.
From a security perspective, if these libraries are open to security scrutiny or certification, they may reduce the opportunity for additional vulnerabilities when implementing the regulatory requirements, and so have a positive impact of \ac{IoT} device security.

\subsection{Future work}
Further research is needed to map the existing \ac{IoT} platform security features to the relevant regulations, to highlight the remaining areas that will need to be addressed.
This could also include a mapping of regulations to any existing software solutions. These existing solutions could be adopted by platform manufacturers to support device engineers when implementing specific regulations.
This paper has only looked at a small selection of  security features from the standards. Further research could look to extend this beyond the baseline security recommendations and also consider how device engineers are supported when needing to implement privacy features.
Research has already looked at the usability of cryptographic \ac{API}s~\cite{acar_comparing_2017}, this research could be extended to the usability of the cryptographic \ac{API}s provided by platform manufacturers and \ac{TEE}s. 
A further study could be conducted with device engineers to understand how they translate the technical documentation found in data sheets to security regulation requirements. Potential research questions could be: 
\begin{itemize}
  \item What issues do device engineers find when translating technical data sheets to security features?
  \item Where do device engineers go to get examples of good security practices?
  \item What could be done to make this translation easier?
\end{itemize}
Researchers have previously looked at the software development process and testing frameworks. This research could be extended to consider the following questions:
\begin{itemize}
  \item How well does \ac{IoT} device development integrate into existing software development processes? 
  \item Do existing testing frameworks extend to \ac{IoT} development? 
\end{itemize}

\section{Conclusions}
We find that platform manufacturers are providing features for device engineers to implement basic \ac{IoT} security features. These security primitives include: \ac{TEE}, the secure boot process, device identity keys, and cryptography functions (RQ1).
The platform manufacturers do not go beyond the security primitives to support device engineers to deliver regulatory requirements such as unique per device passwords.
The level of detail found in the documentation for these features varies depending on the manufacturer. Most manufacturers focus on the hardware functionality, and not how a device engineer can use the functionality to deliver a specific security feature. 
Six of the nine \ac{IoT} platforms reviewed provided some additional support to help device engineers to use these security features correctly (RQ2). These manufacturers provided developer documentation, step-by-step guides or example code.
Some of the sample code included hard coded credentials or defaulted to a less or non-secure implementation.
Three of the platform manufacturers have a \ac{PSA} Certification~\cite{psacertified_psa_2023} for the device we reviewed.
We find no evidence for discussions of \ac{IoT} standards such as ETSI EN 303 645~\cite{etsi_cyber_2020} or regulations such as the \ac{EU} Cyber Resiliency Act (EU-CRA)~\cite{eu_cyber_2023} within the platform manufacturers' technical documentation or developer support material. 
The platform manufacturers take different approaches to support device engineers to take advantage of basic security features (RQ3):
\begin{itemize}
  \item Cloud services deployment
  \item \ac{OS} and security features support
  \item Data-sheets
  \item Limited security features
\end{itemize}
We find that even when the hardware supports a security feature, there is limited guidance for \ac{IoT} device engineers. Device engineers are required either to use the cloud implementation or become a security expert. 
We find a complex standards and regulatory landscape for platform manufacturers and \ac{IoT} device engineers to navigate. 
Platform manufacturers provide no support for other regulatory requirements, such as unique per device passwords.
We conclude that the platform manufacturers, regulators and standards bodies will need to do considerably more to support \ac{IoT} device engineers if we wish to improve the security outcomes of \ac{IoT} devices.

\bibliographystyle{acm}
\bibliography{references,bibliography}
\end{document}